\documentclass[aps,prl,twocolumn,superscriptaddress,showpacs]{revtex4-1}

\usepackage{amsmath}
\usepackage{graphicx}
\usepackage{calc}
\usepackage{bm}

\usepackage[T2A]{fontenc}
\usepackage[cp1251]{inputenc}
\usepackage[english]{babel}

\begin{document}

\title{ Ferromagnetic response of thin NiI$_2$ flakes up to room temperatures}

\author{N.N. Orlova}
\author{A.A. Avakyants}
\author{A.V. Timonina}
\author{N.N. Kolesnikov}
\author{E.V. Deviatov}

\affiliation{Institute of Solid State Physics of the Russian Academy of Sciences, Chernogolovka, Moscow District, 2 Academician Ossipyan str., 142432 Russia}

\date{\today}

\begin{abstract}
We investigate the magnetic response of thin NiI$_2$ flakes for temperatures above 80~K. Since no magnetic ordering is expected for bulk NiI$_2$, we observe clear paramagnetic response for massive NiI$_2$ single crystals. In contrast, thin NiI$_2$ flakes show well-defined ferromagnetic hysteresis loop within $\pm2$~kOe field range.  The value of the response does not scale with the sample mass, ferromagnetic hysteresis can be seen for any flake orientation in the external field, so it originates from the sample surface, possibly, due to the  anisotropic exchange (Kitaev interaction). The observed ferromagnetism is weakly sensitive to temperature up to 300~K. If a flake is multiply exposed to air, ferromagnetic hysteresis is accompanied by the periodic modulation of the magnetization curves, which is usually a fingerprint of the  multiferroic state. While NiI$_2$ flakes can not be considered as multiferroics above 80~K, surface degradation due to the crystallohydrate formation decreases the symmetry of NiI$_2$ surface, which produces the surface ferroelectric polarization in addition to the described above ferromagnetic one.
\end{abstract}


\maketitle

\section{Introduction}

Recent interest to  multiferroics is mostly connected with layered van der Waals materials and van der Waals heterostructures~\cite{reviewMF}. In the simplest case, multiferroic  heterostructure is composed from alternating ferroelectric and ferromagnetic monolayers~\cite{vdW HSs,vdW HSs1}. Multiferroic state has also been  predicted theoretically for layered van der Waals single crystals~\cite{TypeIIMF_DFT}. Experimentally, multiferroic properties have been demonstrated for some dihalides (MX$_2$, X = Cl, Br, I) like CrI$_2$~\cite{CrI2_MF}, MnI$_2$~\cite{MnI2_MF}, CoI$_2$ and NiI$_2$~\cite{NiI2+CoI2_MF,NiI2_MF}.

Among these materials, NiI$_2$ is characterized~\cite{NiI2+CoI2_MF} by  one of the highest transition temperatures $T_{N_2}$= 59.5~K. It is also known as type-II multiferroic~\cite{TypeIIMF_DFT}, where ferroelectricity can only appear in the specific magnetically ordered   state~\cite{classificMF}. For NiI$_2$, the structural transformation to monoclinic noncentrosymmetric lattice is accompanied by transition to a helimagnetic state that displays  finite electric polarization~\cite{NiI2+CoI2_MF, NiI2_MF, TN1_TN2,TN1_TN2_1}. 

It is surprising that multiferroic state can survive down to monolayers~\cite{optical_exp}. Electromagnetic multiferroicity was confirmed in few layers and monolayers of NiI$_2$, which makes it the first established two-dimensional multiferroic~\cite{NiI2_MF, optical_exp}. While decreasing the crystal thickness to monolayers, symmetry requires appropriate changes in the ground state and, therefore, the transition temperature $T_{N2}$. To explain the helical ground state  in the monolayer NiI$_2$, Kitaev interaction~\cite{NI2_ML_spinTex,kitaev2,kitaev3} and a biquadratic term~\cite{biquadratic} have recently been proposed~\cite{Kitaev_iter}.  For monolayer samples, $T_{N2}$ is found~\cite{NiI2_MF,optical_exp} to be decreased to 21~K.   The transition temperature is monotonically increasing with number of layers, so it is  41~K for the four-layer samples~\cite{optical_exp,NiI2_MF}, 58~K for the 60-layer ones~\cite{NiI2_R(T)}, which is close to the bulk value~\cite{NiI2+CoI2_MF} $T_{N_2}$= 59.5~K.  Thus, a 100-layer sample should be regarded as a massive single crystal for its multiferroic properties.   

While the multiferroic state appears below $T_{N2}$, the  magnetic ordering is known even at higher temperatures~\cite{NiI2+CoI2_MF,NiI2_MF}. Bulk NiI$_2$ shows~\cite{ordering} interlayer antiferromagnetic and intralayer ferromagnetic orders below $T_{N1}$=76~K. In contrast to $T_{N2}$, the ordering temperature $T_{N1}$ is predicted~\cite{TN1_ML} to increase in monolayers up to 178~K, while the type of magnetic ordering is still debatable~\cite{magnetic-state,state2}. In general, Kitaev exchange term, when combined with magnetic frustration, can lead to an emergent chiral interaction, which is also responsible for topological spin structures~\cite{NI2_ML_spinTex}.  

Experimental investigation of the magnetic response for NiI$_2$ monolayers is seriously complicated by known NiI$_2$ degradation due to the crystallohydrate formation~\cite{NiI2_R(T)}. On the other hand, theoretical consideration on symmetry and Kitaev interaction should be also valid for  NiI$_2$ surface. The surface-defined magnetic response can be dominant for the thin flakes, which are more accessible for direct magnetic investigations than monolayers. 

Here, we investigate the magnetic response of thin NiI$_2$ flakes for temperatures above 80~K. Since no magnetic ordering is expected for bulk NiI$_2$, we observe clear paramagnetic response for massive NiI$_2$ single crystals. In contrast, thin NiI$_2$ flakes show well-defined ferromagnetic hysteresis loop within $\pm2$~kOe field range.  The value of the response does not scale with the sample mass, ferromagnetic hysteresis can be seen for any flake orientation in the external field, so it originates from the sample surface, possibly, due to the  anisotropic exchange (Kitaev interaction).

\section{Samples and techniques}

\begin{figure}
\includegraphics[width=1\columnwidth]{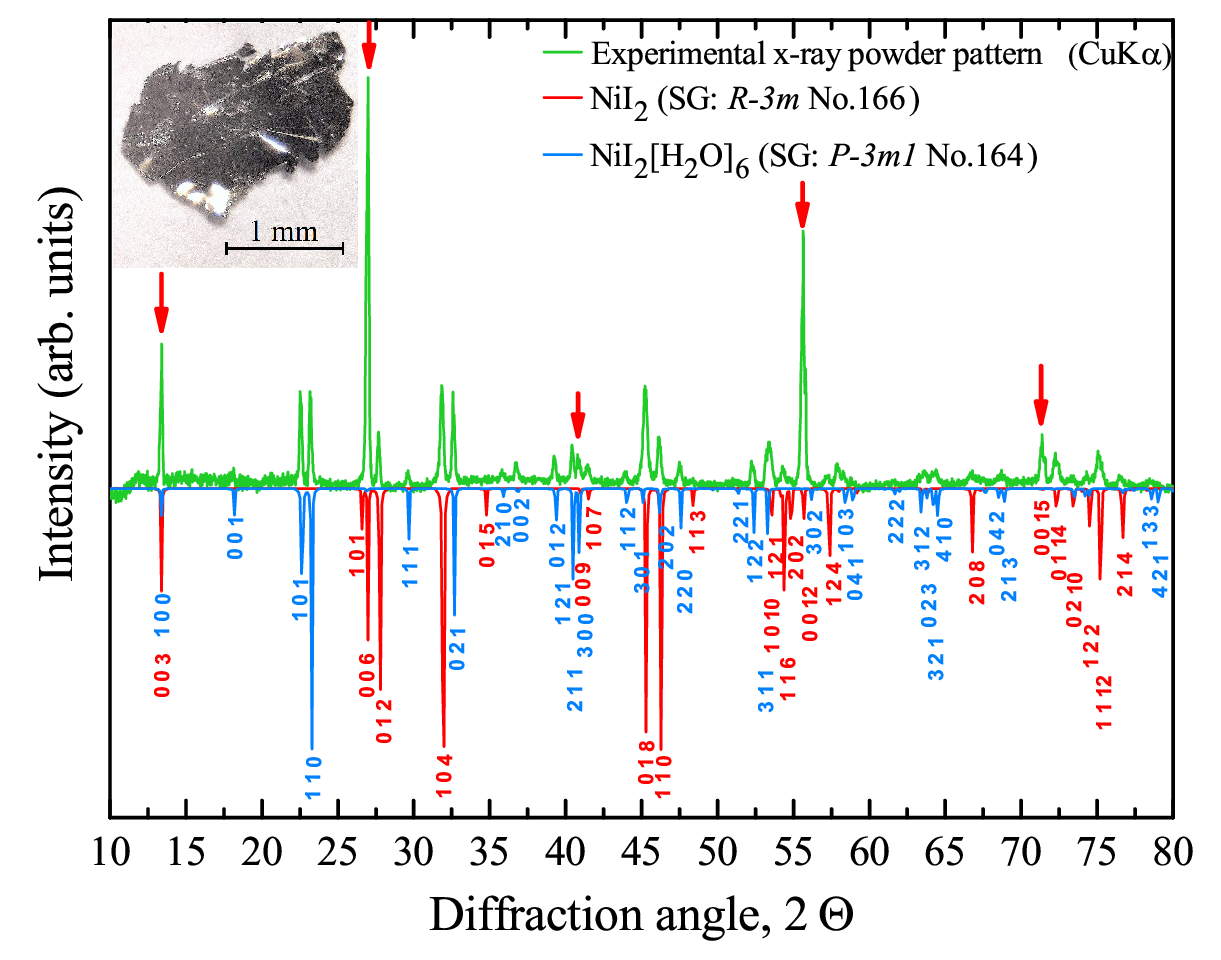}
\caption{(Color online) The powder X-ray diffraction analysis of NiI$_2$ single crystals. The experimental X-ray powder pattern is shown by the green curve. The model X-ray diffraction patterns of pure NiI$_2$ and  NiI$_2$[H$_2$O]$_6$ phases are shown by red and blue curves, respectively. Analysis confirms the known NiI$_2$ structure with trigonal syngony R$\bar{3}$m (space group No.166, CdCl$_2$ prototype) with some admixture of nickel iodide hexahydrate phase, see the text. The data are obtained at room temperature. Intensity of the diffraction peaks shows distinct texture due to the (001) plane (red arrows), which well corresponds to  the layered NiI$_2$ single crystal structure. Inset: optical image of a large single crystal flake, which also shows NiI$_2$ layers. 
}
\label{x-ray}
\end{figure}

NiI$_2$ single crystals were grown by iodine transport in the evacuated silica ampule. The initial load consisted of the mixture of nickel (15x1x0.5~mm$^3$ nickel foil stripes, 99.9\%)  and iodine (99.5\%), taken in the stoichiometric ratio. The ampule was placed in the two-zone furnace,  the load zone was kept at 700$^{\circ}$C while the growth zone was cooled to 550$^{\circ}$C. The distance between the zones was 150 mm. The obtained NiI$_2$ single crystals  are of layered structure, see the image  in the  inset to Fig.~\ref{x-ray}. NiI$_2$ composition and structure are verified by energy-dispersive X-ray spectroscopy and X-ray diffraction analysis, respectively. 

NiI$_2$ is characterized by degradation in ambient conditions~\cite{NiI2_R(T)} due to the crystallohydrate formation. In our experience, 20~$\mu$m thick flake obtains characteristic cyan color~\cite{hydrate} for about 5 hours in air. However, crystallohydrate appears on the flake surface even for the short air depositions, e.g. while the flake is transferred to the sample holder, see the X-ray diffraction pattern in Fig.~\ref{x-ray}. 

The powder X-ray diffraction analysis confirms the main NiI$_2$ phase with some admixture of nickel iodide hexahydrate phase, as depicted in Fig.~\ref{x-ray}.  The known NiI$_2$ structure with trigonal syngony R$\bar{3}$m (space group No.166, CdCl$_2$ prototype) is confirmed. Intensity of the diffraction peaks shows distinct texture due to the (001) plane, which well corresponds to the layered NiI$_2$ single crystal structure. The data are obtained at room temperature, while this NiI$_2$ structure is known to survive~\cite{TN1_TN2,TN1_TN2_1} until 59.5~K. Below 59.5~K, the structure is monoclinic, due to a slight distortion from the C-centered lattice~\cite{struct1}. Thus, the performed X-ray diffraction analysis certainly valid in our temperature range 80--300~K. 

Since we can not avoid surface degradation completely, we should control it by definite	 sample handling. The initial bulk NiI$_2$ single crystal is stored in vacuum in the sealed ampule. After opening to air, an exfoliated flake is immediately placed in the  nitrogen flow cryostat and cooled down to 80~K. The residual part of the initial NiI$_2$ single crystal is stored in liquid nitrogen. The second sample is obtained by warming it to the room temperature in the flow of dry nitrogen, mechanical exfoliation of a thin flake, and cooling down the NiI$_2$ single crystal again. Despite the all precautions, the next sample is always more corrupted than the previous one, which allows to control possible crystallohydrate effects.
 
 We investigate magnetic response of thin NiI$_2$ flakes by Lake Shore Cryotronics 8604 VSM magnetometer equipped with nitrogen flow cryostat. A flake is mounted to the magnetometer sample holder by  low temperature grease, which has been tested to have a negligible magnetic response. The flake's surface can be rotated in magnetic field both for the side mount case in Figs.~\ref{NiI2(120K)},~\ref{NiI2_M(T)1},~\ref{NiI2_M(T)2} (a),  and for the top mount orientation in Fig.~\ref{NiI2_M(T)2} (b). In the every case, we perform centering and saddling procedures to establish the correct sample position in the magnetometer.  

We investigate sample magnetization by standard method of the magnetic field gradual sweeping between two opposite saturation values to obtain hysteresis loops at different temperatures. Apart from the  hysteresis measurements, we perform first order reversal curve (FORC) analysis~\cite{FORCanalysis,Hr}, which is of growing popularity nowadays. The raw FORC data are obtained as multiple magnetization $M(H,H_r)$ curves~\cite{FORCanalysis,Hr}. Before every curve, the magnetization is stabilized at fixed positive saturation field $H_s$. As a second step, the field is changed to the chosen reversal field $H_r$, so the $M(H)$ curve can be recorded toward the positive saturation field $H_s$. For the next FORC curve, the starting reversal field $H_r$ is shifted to the lower magnetic field.  The FORC density distribution $\rho = -1/2 (\partial^2 M(H,H_r) / \partial H \partial H_r)$  is calculated by standard Lake Shore Cryotronics software. Here,   $H_u = \frac{1}{2}(H + H_r)$ is known as the interaction field and $H_c = \frac{1}{2}(H - H_r)$ is the coercive field. FORC analysis provides information on the magnetization reversal, which can not be obtained from standard  hysteresis loops~\cite{FORCtheory,FORCtheory1}.

\section{Experimental results}

\begin{figure}[t]
\center{\includegraphics[width=\columnwidth]{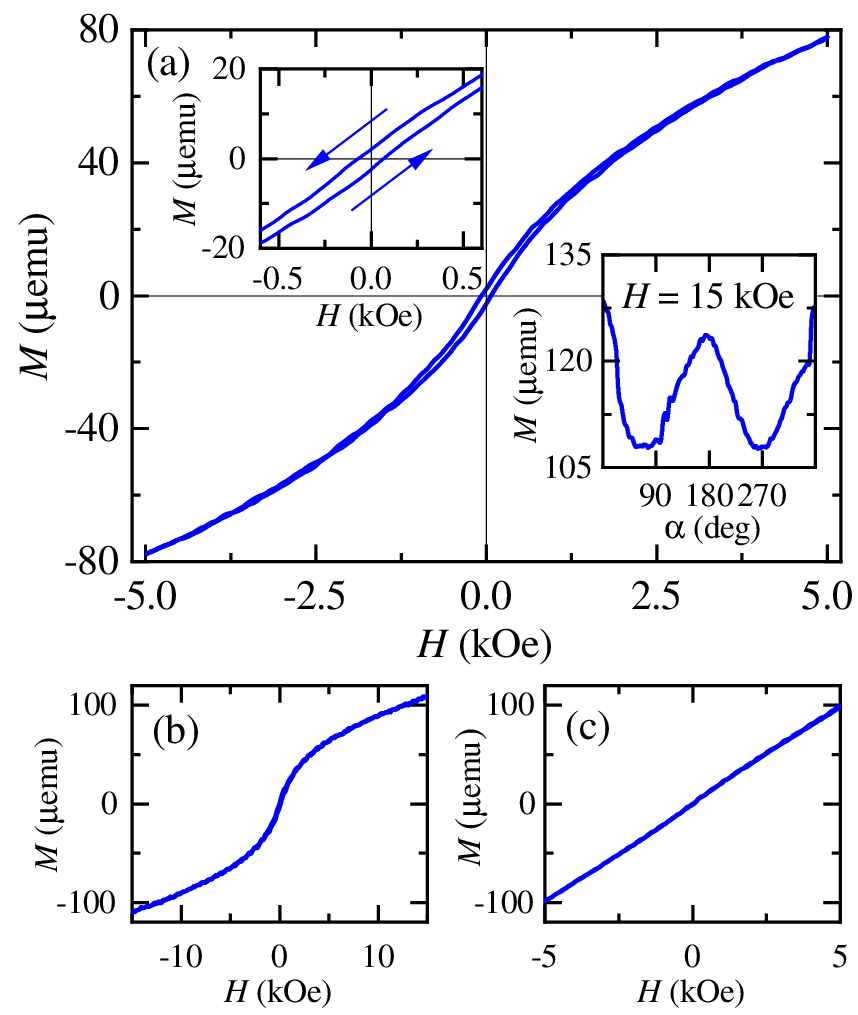}}
\caption{(Color online) Magnetization behavior of NiI$_2$ flakes for the side mount orientation, $\alpha$ is the angle between the flake surface and the magnetic field. The data are obtained at 120~K temperature. (a) Ferromagnetic hysteresis for the  18~$\mu$m thick, 0.23~mg mass flake, magnetic field is parallel to the flake surface, $\alpha=0$. The hysteresis is observed within $\pm$2~kOe field range. For clarity, the central region is enlarged in the left inset. Right inset shows typical ferromagnetic anisotropy of magnetization in high fields (15~kOe), $M(H)$ is diminished for normal field orientation, $\alpha=90^{\circ}$. (b) $M(H)$ curves in a wide field range, $\alpha=90^{\circ}$. Outside the hysteresis region $\pm$2~kOe, $M(H)$ shows linear paramagnetic dependence. (c) Magnetization curves for a massive  sample (100~$\mu$m,  0.88~mg). The linear dependence shows clear paramagnetic response, as it should be expected~\cite{ordering,TN1_ML} for bulk NiI$_2$ at temperatures above $T_{N1}=76$~K. }
\label{NiI2(120K)}
\end{figure}

Fig.~\ref{NiI2(120K)} (a) shows typical magnetization behavior of NiI$_2$ thin flakes. To our surprise, 18~$\mu$m thick, 0.23~mg mass flake demonstrates clear ferromagnetic hysteresis even at 120~K, i.e. well above the bulk $T_{N1}$=76~K. We also observe typical ferromagnetic anisotropy of magnetization, see  the right inset to Fig.~\ref{NiI2(120K)} (a). This anomalous hysteresis is enlarged in the left inset to Fig.~\ref{NiI2(120K)} (a), it appears within $\pm$2~kOe field range. In higher fields,   $M(H)$ is of the linear paramagnetic behavior, as depicted in Fig.~\ref{NiI2(120K)} (b). 

In contrast, massive samples show clear paramagnetic response, as it should be expected from previous publications~\cite{ordering,TN1_ML}, see Fig.~\ref{NiI2(120K)} (c) for the 100~$\mu$m thick,  0.88~mg mass flake. It is important that $M(H)$ values at 15~kOe does not scale with the sample mass in Fig.~\ref{NiI2(120K)} (a) and (c), so the ferromagnetic response might be attributed to the flake surface. 

\begin{figure}[t]
\center{\includegraphics[width=\columnwidth]{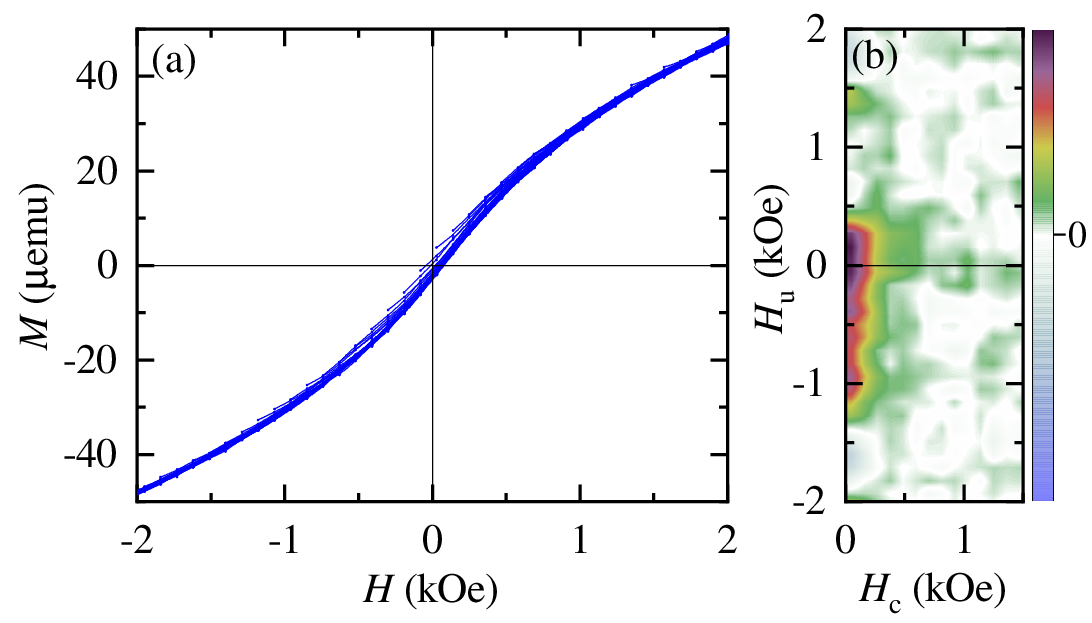}}
\caption{(Color online) FORC analysis for the same 18~$\mu$m thick, 0.23~mg mass flake as in Fig.~\ref{NiI2(120K)}. (a) Raw FORC data, obtained as multiple magnetization $M(H,H_r)$ curves~\cite{FORCanalysis,Hr}. Before every curve, the magnetization is stabilized at fixed positive saturation field $H_s=2$~kOe. The starting reversal field $H_r$ is gradually shifted to the lower fields. The data are obtained at 120~K. Magnetic field is parallel to the flake surface, $\alpha=0$. (b) The calculated FORC density diagram, where  $H_u = \frac{1}{2}(H + H_r)$ is known as the interaction field and $H_c = \frac{1}{2}(H - H_r)$ is the coercive field. Single peak in $\rho(H_u,H_c)$ and the so-called open contours at the $H_u$ axis are usually regarded as a fingerprint of the multidomain ferromagnet~\cite{FORCtheory,FORCtheory1}.}
\label{NiI2(FORC120K)}
\end{figure}

The conclusion on the ferromagnetic behavior of NiI$_2$ thin flakes is supported by FORC analysis in  Fig.~\ref{NiI2(FORC120K)}. The raw FORC curves and the calculated FORC density diagram $\rho(H_u,H_c)$ are shown in (a) and (b), respectively, at 120~K temperature for the  sample from Fig.~\ref{NiI2(120K)} (a). The raw magnetization $M(H,H_r)$ curves confirm ferromagnetic response of the sample. We also observe single peak in $\rho(H_u,H_c)$, which is centered at low $H_c$ values with the so-called open contours at the $H_u$ axis. This behavior is usually regarded as a fingerprint of the multidomain regime for a ferromagnet~\cite{FORCtheory,FORCtheory1}. The peak center is slightly shifted to the positive values of the interaction field $H_u$, which corresponds to the dipolar interaction between domains~\cite{FORCtheory,FORCtheory1}.

\begin{figure}[t]
\center{\includegraphics[width=\columnwidth]{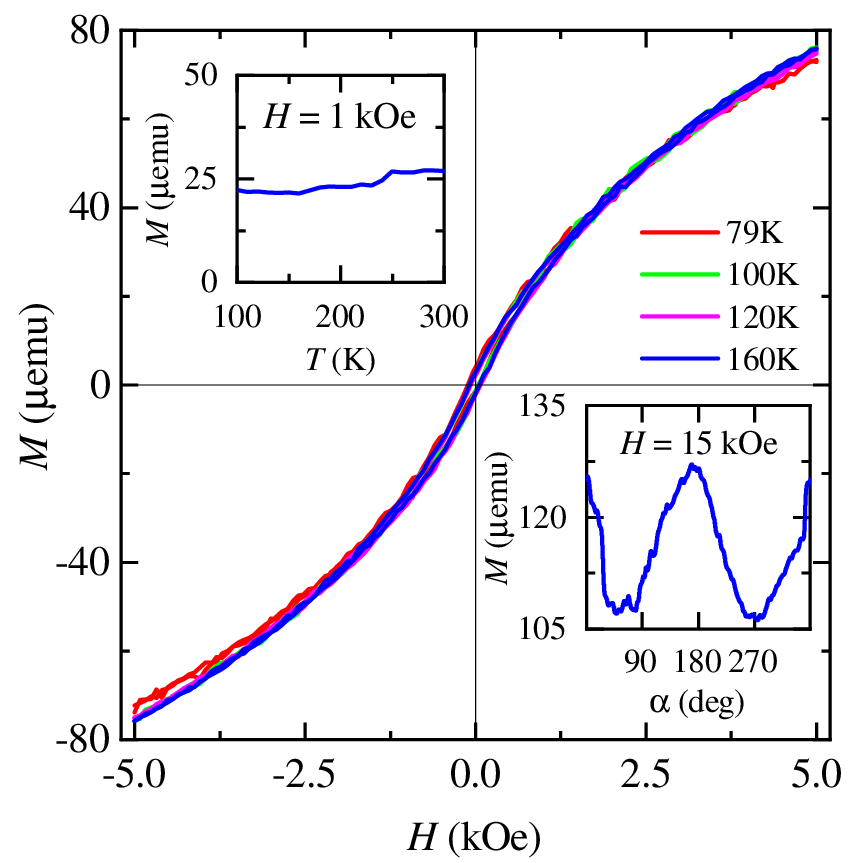}}
\caption{ (Color online) Magnetization curves at 79~K, 100~K, 120~K, 160~K temperatures for the same 18~$\mu$m thick, 0.23~mg mass flake as in Fig.~\ref{NiI2(120K)}. $M(T)$ dependence is weak  up to the room temperature, as it is shown in the left inset. $M(\alpha)$ is shown in the right inset at 160~K temperature for 15~kOe field, it coincides well with the result at 120~K in  Fig.~\ref{NiI2(120K)} (a).}
\label{NiI2_M(T)1}
\end{figure}

Ferromagnetic response of NiI$_2$ thin flakes can be demonstrated at different temperatures, see Fig.~\ref{NiI2_M(T)1}. The magnetization curves coincide well from 79~K to 160~K, as shown in the main field of   Fig.~\ref{NiI2_M(T)1}. Moreover, the magnetization $M$ value is nearly constant at 1~kOe up to the room temperature, see the left inset to Fig.~\ref{NiI2_M(T)1}.  The angle dependence at 160~K is also identical to one at 120~K, cp. the insets to Figs.~\ref{NiI2(120K)} (a) and Fig.~\ref{NiI2_M(T)1}, so there is no sizable temperature dependence for any angle $\alpha$ between the flake surface and the magnetic field for the side-mount orientation of the sample.

\begin{figure}[t]
\center{\includegraphics[width=\columnwidth]{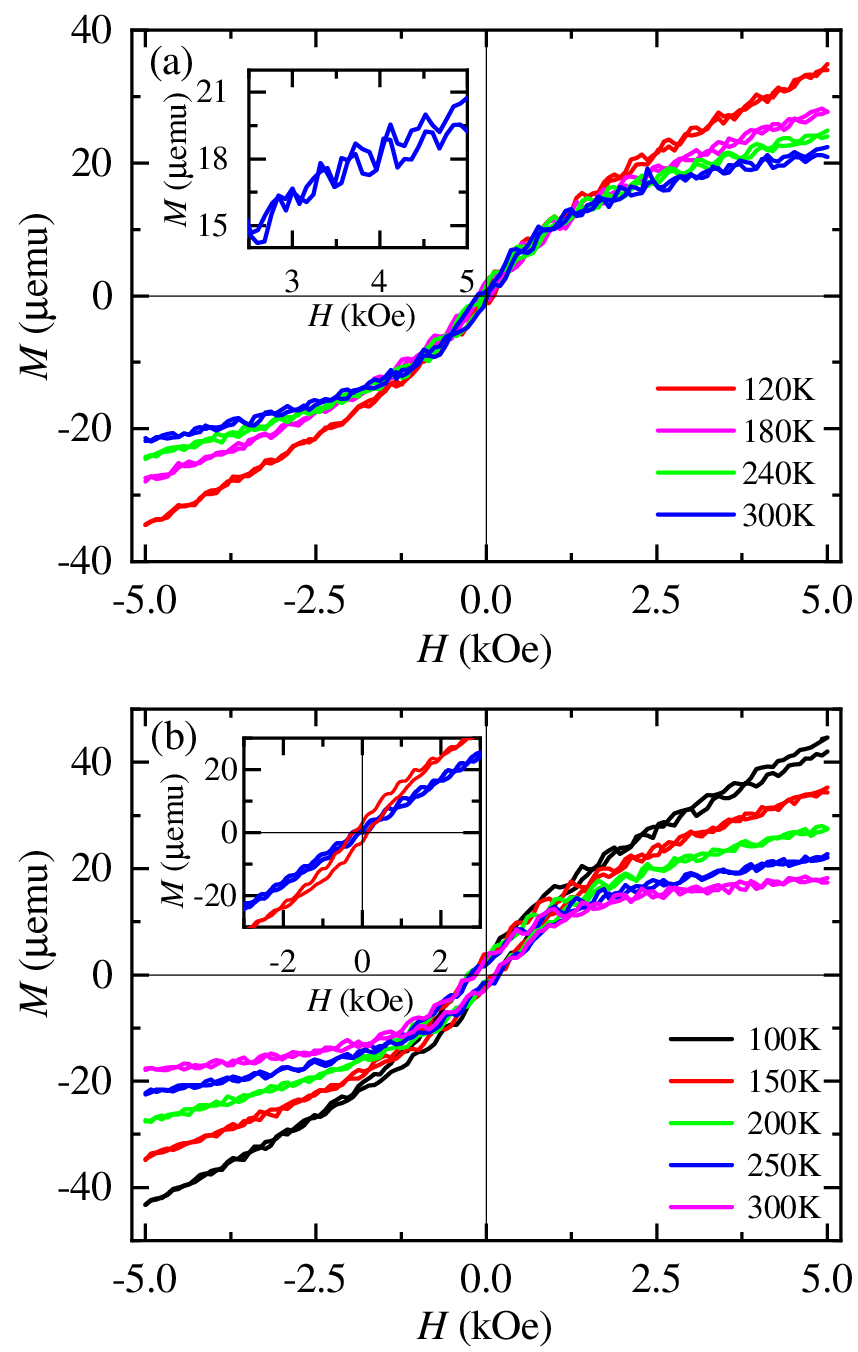}}
\caption{ (Color online) Ferromagnetic hysteresis for a similar (15~$\mu$m,  0.20~mg) NiI$_2$ flake, which is exfoliated after several warmings of the initial crystal to room temperature. (a) Side mount orientation of the sample, $M(H)$ curves are shown at 120~K, 180~K, 240~K, 300~K temperatures. The hysteresis  appears in the same field  range $\pm$2~kOe, so coersitivity is independent on the particular sample and surface degradation.  However, the saturation level is diminished in this case.  As an additional result of surface hexahydrate, shallow oscillations appear as modulation of $M(H)$. The oscillations are perfectly reproducible as it is demonstrated in the left inset. (b) Similar $M(H)$ data for the top mount sample orientation at 100~K, 150~K, 200~K, 300~K temperatures. Magnetic field is within the flake's plane, there is angle dependence of the ferromagnetic hysteresis, as it is shown in the inset for two $120^{\circ}$ shifted $\alpha$ values.  
 }
\label{NiI2_M(T)2}
\end{figure}

We observe qualitatively similar results for different samples, which also allows to highlight an effect of surface degradation due to the crystallohydrate formation. Fig.~\ref{NiI2_M(T)2} shows  magnetization curves for  a similar (15~$\mu$m,  0.20~mg) flake, which is obtained from the same initial NiI$_2$  crystal after several warmings to room conditions. Clear ferromagnetic hysteresis can be seen for the side mount and for the top mount flake orientations in (a) and (b), respectively, for the temperatures up to 300~K.  We also demonstrate  typical ferromagnetic anisotropy of magnetization, see  the left inset to Fig.~\ref{NiI2_M(T)2} (b). The hysteresis  appears in the same field  range $\pm$2~kOe, so the coersitivity is independent on the particular sample and on the surface degradation.  However, the saturation level is diminished in this case from 80~$\mu$emu to 40~$\mu$emu for the samples of similar masses (0.23~mg and 0.20~mg, respectively). It seems that the surface crystallohydrate diminishes the $M(H)$ for thin NiI$_2$ flakes.

Surprisingly,  Fig.~\ref{NiI2_M(T)2} also shows clear modulation of the magnetization curves by well-reproducible shallow oscillations. We check that there is no oscillations at fixed magnetic field. Some precursors of this behavior can be also seen at low temperatures for the  sample from  Fig.~\ref{NiI2_M(T)1}, see the curves obtained at 79~K and 100~K. In Fig.~\ref{NiI2_M(T)2}, the oscillations can be seen up to 300~K. The oscillations are stronger for the samples, obtained after multiple openings to air, they  are well-reproducible and, therefore, requires consistent explanation, as well as the ferromagnetic hysteresis itself.

\section{Discussion}

As a result, we observe evident ferromagnetic response from thin NiI$_2$ flakes at temperatures well above the bulk $T_{N1}$=76~K ordering temperature, while the thick massive flakes show clear paramagnetic signal in this temperature range. If a flake is exposed to air, ferromagnetic hysteresis is accompanied by the periodic modulation of the magnetization curves.

First of all, no magnetic order can be expected~\cite{TN1_TN2,TN1_TN2_1, NiI2+CoI2_MF,ordering} above $T_{N1}$=76~K for bulk NiI$_2$, which we confirm by clear paramagnetic response for the thick massive flakes in Fig.~\ref{NiI2(120K)} (c).  We can not completely exclude some crystallohydrate admixture, e.g., while a flake is transferred  to the sample holder. However, the  NiI$_2$  crystallohydrate is known to have paramagnetic response~\cite{hydrate,Hexahydrate2}, so it can not be responsible for the observed ferromagnetic hysteresis. 

Bulk NiI$_2$ is a centrosymmetric magnetic semiconductor, so neither spin-orbit coupling nor the Dzyaloshinskii- Moriya interaction are  allowed by the inversion symmetry of NiI$_2$. Incommensurate spin patterns are also too weak to generate non-negligible Dzyaloshinskii-Moriya interaction~\cite{Kitaev_iter}. 

For thin NiI$_2$ flakes, the situation is more sophisticated due to the  anisotropic exchange (Kitaev interaction). Magnetic interactions between localized spins can be generally modeled by the classical spin Hamiltonian, which include the exchange coupling interaction tensors. The latter is generally decomposed into three contributions~\cite{NI2_ML_spinTex}: the isotropic coupling term, defining the scalar Heisenberg model; the antisymmetric term, which corresponds to the Dzyaloshinskii-Moriya interaction and vanishes in the presence of an inversion center; the anisotropic symmetric term also referred to as a Kitaev term. The latter is expected~\cite{NI2_ML_spinTex,Kitaev_iter} to determine the helical ground state below $T_{N2}$. At higher temperatures (above $T_{N_2}$),  Kitaev interaction changes antiferromagnetic ground state to ferromagnetic one~\cite{TN1_ML}, with strong increase of the ordering temperature $T_{N1}$. As a result, the monolayer NiI$_2$ is a ferromagnetic insulator with the calculated $T_{N1}$ about 200~K. 

One can expect, that the Kitaev interaction becomes also important for the thin NiI$_2$ single crystal flakes, which should be a reason to observe magnetic ordering within 80~K -- 300~K temperature range, much above the bulk $T_{N1}$=76~K. As a possible scenario, the above considerations on the Kitaev interaction should be important for the  NiI$_2$ flake surface. The   topological spin structures are predicted~\cite{NI2_ML_spinTex,kitaev2,kitaev3} for the  NiI$_2$ surface, e.g. as the spontaneous formation of skyrmionic lattice with a unique, well-defined topology and chirality of the spin texture due to the Kitaev interaction~\cite{Kitaev_iter}. The surface-defined magnetic response is more important for the thin flakes, which is a good correspondence with our experimental results. Indeed, there is no ferromagnetic response for the massive thick flake in Fig.~\ref{NiI2(120K)} (c), while the magnetization  saturation level is found to be dependent on the level of surface degradation for the samples of the similar masses, see Figs.~\ref{NiI2_M(T)1} and Fig.~\ref{NiI2_M(T)2}. Also, magnetic anisotropy can be seen for any orientation of NiI$_2$ flakes, which also correlates with the surface-defined effect.

As about shallow oscillations in  Fig.~\ref{NiI2_M(T)2}, they appear in the experiment if a flake is multiply exposed to air, i.e. for noticeable surface crystallohydrate. On the other hand, 
shallow oscillations can be expected~\cite{indusi} in the multiferroic state, see also, e.g., Fig.2 in Ref.~\cite{multi_oscill}. Low-symmetric crystallohydrate thin film produces the surface ferroelectric polarization~\cite{Hexahydrate3} in addition to the described above ferromagnetic properties due to the Kitaev interaction. Thus, the crystallohydrate-affected NiI$_2$  surface can be considered as artificial multiferroic even at high temperatures.
 
Multiferroics are materials that exhibit different coexisting ferroic orders such as ferroelectricity, ferromagnetism, or ferroelasticity. Due to the coupling among the different degrees of freedom leading to these ordered states, the order parameters of one state can be controlled by tuning parameters different from their conjugate variable~\cite{MForigin}. In the conditions of our experiment, variation of the magnetic field leads to appearance of the electric field due to the magnetoelectric coupling. Electric field produces mechanical stress in ferroelectrics, which, subsequently, affects magnetization due to the strong magneto-elastic coupling~\cite{indusi}. As a result, variation of the external magnetic field should produce shallow magnetization oscillations.  

This mechanism is especially important for a wide band gap semiconductor material NiI$_2$, with negligible bulk conductivity even at room temperature~\cite{resistanceNiI2}. Thus, shallow oscillations of magnetization for the crystallohydrate-affected NiI$_2$ thin flakes should be considered as additional confirmation of the surface origin ferromagnetism in NiI$_2$.

\section{Conclusion}
As a conclusion, we investigate the magnetic response of thin NiI$_2$ flakes for temperatures above 80~K. Since no magnetic ordering is expected for bulk NiI$_2$, we observe clear paramagnetic response for massive NiI$_2$ single crystals. In contrast, thin NiI$_2$ flakes show well-defined ferromagnetic hysteresis loop within $\pm2$~kOe field range.  The value of the response does not scale with the sample mass, ferromagnetic hysteresis can be seen for any flake orientation in the external field, so it originates from the sample surface, possibly, due to the  anisotropic exchange (Kitaev interaction). The observed ferromagnetism is weakly sensitive to temperature up to 300~K. If a flake is multiply exposed to air, ferromagnetic hysteresis is accompanied by the periodic modulation of the magnetization curves, which is usually a fingerprint of the  multiferroic state. While NiI$_2$ flakes can not be considered as multiferroics above 80~K, surface degradation due to the crystallohydrate formation decreases the symmetry of NiI$_2$ surface, which produces the surface ferroelectric polarization in addition to the described above ferromagnetic one.

\section{Acknowledgement}

We wish to thank S.V.~Simonov for X-ray sample characterization.  We gratefully acknowledge financial support  by the  Russian Science Foundation, project RSF-23-22-00142, https://rscf.ru/project/23-22-00142/.

\end{document}